Exploring the determinants of Bitcoin's price: an application of Bayesian Structural Time Series

Student: Obryan Poyser

Supervisor: Jordi Perdiguero

June 2017

# Abstract

Currently, there is no consensus on the real properties of Bitcoin. The discussion comprises its use as a speculative or safe haven assets, while other authors argue that the augmented attractiveness could end accomplishing money's functions that economic theory demands. This paper explores the association between Bitcoin's market price and a set of internal and external factors using Bayesian Structural Time Series Approach. I aim to contribute to the discussion by differentiating among several attractiveness sources and employing a method that provides a more flexible analytic framework that decompose each of the components of the time series, apply variable selection, include information on previous studies, and dynamically examine the behavior of the explanatory variables, all in a transparent and tractable setting. The results show that the Bitcoin price is negatively associated with a neutral investor's sentiment, gold's price and Yuan to USD exchange rate, while positively related to stock market index, USD to Euro exchange rate and variated signs among the different countries' search trends. Hence, I find that Bitcoin has mixed properties since still seems to act as a speculative, safe haven and a potential a capital flights instrument.

Short

Currently, there is no consensus on the real properties of Bitcoin. The discussion comprises its use as a speculative or safe haven assets, while other authors argue that the augmented attractiveness could end accomplishing money's functions that economic theory demands. This paper explores the association between Bitcoin's market price and a set of internal and external factors using Bayesian Structural Time Series Approach. The results show that Bitcoin has mixed properties since seems to currently act as a speculative, safe haven asset and a potential a capital flights instrument.

## Figures index

## Table index

# I. Introduction

Digital currencies have been increasing in attention during last years, inevitably it reached in academia, finance, and public policy atmospheres. From the academia perspective, this importance arise on the fact that it has features that generate several conflicts in political and financial environments. Even the definition is ambiguous, as a product of an information of technology conception, it can be defined as a protocol, platform, currency or payment method (Athey et al. 2016). Among digital currencies, Bitcoin has been capturing almost all of its reflection, this virtual currency was created in 2009 and serves as a peer-to-peer version of electronic cash that let to do transactions on the internet without the intermediation of the financial system (Nakamoto 2008).

Digital coins or cryptocurrencies, named in such way due to their characteristic of using encryption systems that regulate the creation of coins and transfers have to be identified from an economic analysis perspective. Hence, it is important to examine which social, financial and macroeconomic factors determine its price in order to know the scope and consequences of the economy.

Bitcoin as well as alternative coins (Altcoin) have been vastly criticized due to its declared avoidance from the financial system that derives from the inability of actions of the government, which makes impossible to levy cash movements, control money laundry and fight illegal activities among other issues. Events such Chinese government's decision to blackout Bitcoin in 2013 (Baek & Elbeck 2014) and the call made in early 2017 by the PBoC to main Chinese Bitcoin exchange firms to discuss illegal activities that might have happened in China. Other events including the bankruptcy announcement of Mt. Gox, one of the heads of Bitcoin trading (Yermack 2013) and recently the gain in legitimacy after the Brexit vote (Halaburda 2016; Bouoiyour & Selmi 2016b) have only increased the need to study deeply digital currency. As Böhme, Christin, Edelman, & Moore (2015) claimed that for an economist it is interesting since it has the possibility to "disrupt existing payment systems and perhaps even monetary systems." Additionally, many authors have argued that Bitcoin, resembles a speculative asset attributable to its high volatility. Another characteristic that BTC shows is a high volatility behavior which is consistent with typical speculative assets movements, an aspect that is been also criticized by many financial spokesmen.

Besides all negative aspects aforementioned, there is also positive signs according to experts. For instance, in the first days of February 2017, there occurred a milestone for cryptocurrencies history, Bitcoin's value surpassed again the $1000 psychological threshold since the events of 2013 when it reached more than $1150 per BTC. Accordingly, some authors have argued that it is possible that BTC is entering a mature phase due to the decrease in the price volatility and an increment in acceptance as a payment method in different businesses. In summary, Bitcoin has been through government bans, hacks, and bad reputation, conversely, notwithstanding all these trials, it is still growing and being the most established cryptocurrency of the market. This behavior is generating among users and investors a resilience perception around BTC that might be associated with an increasing the confidence in its future.

The goal of this paper is to explore the implication of search trends as a proxy for interest, macro-financial and internal factors that impact the price. One the main contributions of this study are that it will follow the structural time series approach that allows disaggregating the series in their different components, moreover, they have the ability to let the coefficients vary over time permitting the estimation of sensitivities of different factors or structural changes due to significant events. None of the existent research has accounted for the differentiation of the elements that have dealt with the price dynamics over the time and the possibility that this digital coin is entering a different stage. Moreover, the condition that the search trend and magnitude vary greatly across countries has not been accounted yet, in that sense, this study innovates by applying data-driven methods to specify which groups of search trends have a relevant relationship with Bitcoin price.

This paper is organized as follows. Section one provides an introduction to the case, section two describes the background of Bitcoin, while in section three it is going to be revised some of the most significant literature about the price formation, estimation, political and financial influence and of Bitcoin. Section four shows the nuances of the data that it is going to be used to estimate Bitcoin price measured by the exchange rate with the USD. In the fifth section, the methodology will be explained putting noteworthy emphasis on the state space method. Section six shows the main results regarding the prediction, whereas in section seven there is a discussion about the effect of each set of variables. Finally, the eitgh section defines the core conclusions of the study.

## II. Background

The foundations of digital currencies rely on cryptography advances. The capacity to secure communications drove many researchers to create digital currencies, however, they failed due to their centralization, precious metal backing, counterfeiting and double-spend issues (Antonopoulos 2014). The first problem arose from the characteristic of being settled in specific spaces that were the reason prior digital currencies were susceptible to government prohibition and hackers attacks. Regarding trustiness, it is easy to realize that it might be a problem to prove the authenticity of digital coins, while the issue of preserving the property rights of a set of coins it was also a great inconvenient. The two aforementioned problems were mostly solved by creating digital signatures under an appropriate technological architecture. Particularly, Bitcoin is the ultimate creation after decades of developments and technological applications of cryptography.

Bitcoin is an open source computer program that was invented by an entity under the pseudonym of Satoshi Nakamoto in 2008. According to Antonopoulos (2014), Bitcoin is a set of technologies that established the framework to interchange money named bitcoins (lower case). In detail, it firstly consists of a decentralized peer-to-peer network, this implies that there is no intervention of the government nor financial system, instead, it is a self-organized interconnected set of nodes, where each node represents a buyer or a seller, and only both parts are involved in the transaction (Nakamoto 2008). Secondly, Blockchain serves as the public ledger for all transactions, where in it a set the rules how to create, distribute, interchange, and validate the flow of block of transactions are stablished. Thirdly, the bitcoin is the inherent currency that has the function to represent value and serve as a reward to the operators in the network for securing the distributed ledger (Franco 2014). Regarding this issue, Nakamoto (2008) outlined the rules that determine the amount of "coins" produced over time and the method to create them. There is a determinist rule that specified that the limit of bitcoins will be 21 million bitcoin in the year 2140.

Finally, it is precise to mention that the asymptotic limit of 21 million[1] Bitcoins derives from the *issuance of new bitcoins to reward operators* (miners) *in the network for securing the distributed ledger* (Franco 2014). Miners are individuals with high computational power used to solve algorithm that maintain the network by organizing transaction into blocks. In exchange, their work they receive a fee that will depend on current market activity. The former compensation is 50 bitcoins and this number is halved every four years, and by extension, there will be nearly 210,000[2] blocks for each set of four years.

## III.  Related works

The conventional economic approach to outline what is considered money is based on a set of basic functions. The first function is the medium of exchange, that is, an intermediary mechanism that aligns the demands for each pair of agents present in a trade event. The second function is the ability to work as a unit of measurement, needed to set comparability between the goods and services that are being traded through the interchange. Finally, the third function ability is to store the value over time. Several authors have been trying to interpret the role of bitcoin under the contemporary definition of money, for instance, Bjerg (2016) compared bitcoin to a set of ideal typical theories of money. The author developed the analysis under the principle that bitcoin is "*a commodity money without gold, fiat money without state, and credit money without debt*", and claimed that even though bitcoin is no gold, state or debt backing, it is a mistake to settle for a counterfeit money. On the other side, Yermack (2013) argued that bitcoin does not have the possibility to meet the classical properties of a money since it lacks intrinsic value, long verification process of the transactions and high volatility.

Although the heterogeneity of criteria of the author's bitcoin's adequacy as a currency, almost all of them coincide in the fact that bitcoin future as a currency is mostly linked to the credibility and acceptance from users and merchants (Luther & Salter 2015).

---

[1] This limit can be depicted as a geometric series, and it is straightforward to find that the common factor is 0.5, thus we can calculate the maximum amount of bitcoins by:
$$S_n = \frac{a(1-r^n)}{(1-r)} = 210,000 * \frac{50(1-0.5^\infty)}{1-0.5} \sim 21x10^6$$
[2] This value is easily proved as there is a block each 10 minutes, hence, 144 per day, and given that there are 1460 days in four years, the result is 210.240 blocks.

Further uses might end turning bitcoin into illicit activities platform or as a speculative asset (Bjerg, 2016; Ciaian, Rajcaniova, & Kancs, 2016; Yermack, 2013). Henceforward, the purpose of this study is to analyze bitcoin's price drivers in a dynamic scope in order to shed light on the evolution. Most of the empirical literature addresses bitcoin's price estimation by using social information, financial and macroeconomic variables, however, none of them has been considering time variance of these relationships. Other debate around bitcoin's digital currency and its appropriateness as money can be found in Böhme et al. (2015), Glaser, Zimmermann, Haferkorn, Weber, & Siering (2014), Rogojanu & Badea (2014), Simser (2015) and Wisniewska (2015).

In order to estimate bitcoin's price drivers there have been two main branches of explanatory variables use: those papers which only include sentiment analysis (pure adoption and attractiveness) and other that employed macroeconomic and financial variables, however, most the latter group have also included at least a proxy for investor's attractiveness. Within the first branch, Kaminski (2014) studied how emotions in Twitter influence digital currency market and argued that those sentiments have a moderate correlation with Bitcoin closing price and volume. In extension, Granger causality analysis found that there is no statistical significance for Twitter signals as a predictor, in contrast, this social media an emotional reflection of Bitcoin's price movements. Similarly, Yelowitz and Wilson (2015) collected Google Trends data and anecdotal evidence regarding Bitcoin users to examine the determinants of interest in Bitcoin. According to this paper, computer programming enthusiasts and unobserved illegal activities drive interest of Bitcoin, while political and financial variables effect are less supported. Finally, another contribution was done by Kim et al. (2016) who analyzed social activity in cryptocurrencies communities and constructed a sentiment analysis index in order to explain if those variables affect Bitcoin, Ethereum, and Ripple cryptocurrencies price, finding that the proposed approach predicted variability in the price of low-cost cryptocurrencies.

Regarding the second branch, Garcia, Tessone, Mavrodiev, & Perony (2014) and Kristoufek (2015) have been two of the most influential studies, they addressed the analysis by differentiating between internal and external drivers of Bitcoin's price. Specifically, the latter paper provided a framework to categorize the drivers that might influence Bitcoin's price, additionally, it formalized the role of Bitcoin as a potential hedge or safe-haven asset and described the great influence of China market on it. By applying wavelet coherence method, the author examined potential drivers, such as economic, transactional, technical and interest. Kristoufek opened the discussion of the duality property of Bitcoin (digital currency or speculative asset) by arguing that "*although the Bitcoin is usually considered a purely speculative asset, we find that standard fundamental factors —usage in trade, money supply, and price level—play a role in Bitcoin price over the long term.*" This argument reinforces the idea that it is not all lost for now, however, the author also mentioned that for now it a unique asset that goes from purely financial to speculative.

Recently, by comparing Bitcoin with precious metals, analyzing volatility and adoption, the interest has been extending the understanding about Bitcoins price movements. For instance, Cheah & Fry (2015) found that Bitcoin is prone to substantial speculative bubbles, a result that confirmed Baek & Elbeck (2014) findings, however, the latter paper specified that bitcoin's importance is growing, they expect to become more stable in the future. Georgoula, Pournarakis, Bilanakos, Sotiropoulos & Giaglis (2015) who applied time series analysis to study the impact of economic, technological and Twitter sentiment indicators on bitcoin. According to their results, in the short run, positive Twitter sentiment, Wikipedia search queries, and hash rate have a positive relationship with bitcoin's price, while USD to Euro exchange rate a negative one. Through the employment of a VEC model, they found that in the long run bitcoin's price is positively related with bitcoins in circulation and negatively associated with S&P500 index. Other worthy results can be seen in Bouoiyour & Selmi (2016a), (2016b); Bouri, Gupta, Tiwari, & Roubaud (2017); Ciaian et al. (2016a); Ciaian, Rajcaniova, & Kancs (2016b) and Dyhrberg (2016).

As it has been shown before, most of the empirical work rely on Google trends and Twitter sentiment as a measure of attractiveness, however, I have found that the behavior of the search queries is not homogeneous across countries not static over time. From an empirical perspective, besides distinguishing about countries particularities, it is also important to the include a indicator of illegal activities and more importantly, the effect of relevant political events (Cheah & Fry, 2015 provided a similar approach to study speculative bubbles), that might have altered the level, drift or slope of the series. The disaggregated investor's attractiveness and the testing for influential geopolitical events are the main novelty for the empirical literature variable inclusion. The next two sections will provide an explanation of the descriptors of the model.

1. Price drivers

Following the framework proposed by (Ciaian, Rajcaniova & D. Kancs 2016; Ciaian, Rajcaniova & dArtis A. Kancs 2016; Kristoufek 2015) I will differentiate between three types of drivers organized into internal and external factors. By internal, it tries to find the supply and demand variables that are directly derived from information of Bitcoin platform. On the other side, external factors will be composed by attractiveness and macro-financial drivers (figure 1).

Figure 1: Bitcoin price drivers

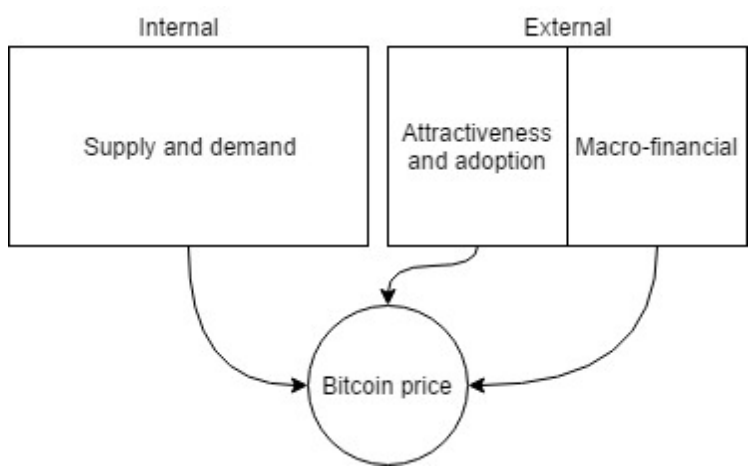

Internal factors

Bitcoin has a controlled supply of coins settle by block height and block reward values, this is intrinsically related to the mining process.

From this situation we can imply two things, firstly, bitcoin's supply is exogenously determined and secondly, it is deflationary constructed[3]. The aforementioned problem was discussed by (Yermack 2013; Böhme et al. 2015; Garcia et al. 2014), and the consensus is that it represents a serious drawback in its way to becoming a real currency according to the economic principles.

---

[3] Many economist have stressed about the deflationary spiral that bitcoin represents to an economy. See (Hanley 2013; Barber et al. 2012; Grinberg 2011) for a broader discussion.

Given that supply is deterministic, only the demand side that can affect bitcoin's price (Ciaian, Rajcaniova & D. Kancs 2016; Kristoufek 2015; Baek & Elbeck 2014). Between the internal variables we have: bitcoins in circulation, transaction volume, hash rate and mining difficulty, these variables will be explained in detail in the data section.

External factors

Besides pure demand variables, other forces might influence bitcoin's price, for instance, some authors have been studying the role of Bitcoin as a safe haven and hedge[4] instruments. Bouoiyour & Selmi (2016a) examined the interconnection of precious metals and bitcoin with volatility in financial markets. They found that gold, silver and Bitcoin appropriateness as a hedge and safe haven is not constant over time, but particularly, Bitcoin acts as a weak safe-haven in the short run, and as a hedge in the long run. Likewise, in a previous study Kristoufek (2015) found one period of time that showed correspondence amongst the Financial Stress Index and bitcoin's price. Nonetheless, in a recent paper Bouri, Gupta, Tiwari, & Roubaud (2017) backed Bouoiyour & Selmi results, since by studying whether bitcoins can be a hedge or safe haven asset under market uncertainty scenarios, the authors found that Bitcoin acts as a hedge since it reacted positively to great (low and up) financial movements, especially in the short run. From this results, it is important to study bitcoin's relationship with financial indicators and precious metals prices in a dynamic environment.

As it was previously stated, it is difficult to define Bitcoin since it has several capabilities, nonetheless, the payment method and investing asset is predominantly addressed by an attractiveness proxy, this study will include such variables. In most cases authors have rely on Google search trends and Wikipedia articles views (Kristoufek 2015; Glaser et al. 2014), Twitter sentiment analysis (Kaminski 2014; Georgoula et al. 2015) and online communities reactions (Ciaian, Rajcaniova & D. Kancs 2016; Dwyer 2015; Kim et al. 2016). Among all the variables in the studies, the attraction has the most relevant variance explanation power. Nevertheless, none of the previous works have realized that trends are not uniform across countries, that is, search trends in the Unite States are significantly different from China's, as far as I know, this has not been accounted in the empirical literature. On this matter, this paper provides an innovation in comparison with other studies, the further description can be found in posterior sections.

---

[4] Increasing risks in financial markets have established the need to invest in other type of assets, historically they have been precious metals. The theoretical argument about the existence of such assets is that investors have incentives to reduce losses in times of market stress. According to (Baur & Lucey 2010) we can distinguish between three types these assets, hedge, diversifier and safe-haven. Hence, a safe-haven and hedge are defined as an asset that is uncorrelated or negatively correlated with another asset, with the distinction that the former behaves as it in those market situations with high degree of stress and turmoil. Finally, a diversifier is positively correlated with another asset. For a broader discussion on gold's and other asset application see (Baur & Lucey 2010; Baur & McDermott 2010; Ciner et al. 2013).

The Internet is set to be disruptive since it has the capability to change the way we interpret and behave. Bitcoin's collection of properties technologies can dramatically change our economy as well. Likewise, it is important to underline that given that the conceptual foundation of digital currencies was not product the economic thinking, rather than information technology area, a lagged interest was given to the research about the definition and scope from the sight of economics scene. Recently, many authors have been studying the impact that Bitcoin as a currency has on the financial market and its relation to fiat money. As Franco (2014) argues, whether the value of bitcoin has the economic future relies on the forces driven by its application. In the next section, I will provide some of the most relevant the economic literature on bitcoin, while other conceptual and technical description of the technology behind can be revised in (Antonopoulos 2014; Franco 2014; Nakamoto 2008; Bonneau et al. 2015).

## IV. Data

As it was mentioned before, I will study the Bitcoin exchange rate with the USD (price) as target variable. The website [www.blockchain.info](www.blockchain.info) is a wallet and block information service and it is the main source for internal factor variables. This site continuously records information about the BTC/USD on daily frequency (figure 2), additionally, it provides data about other variables.

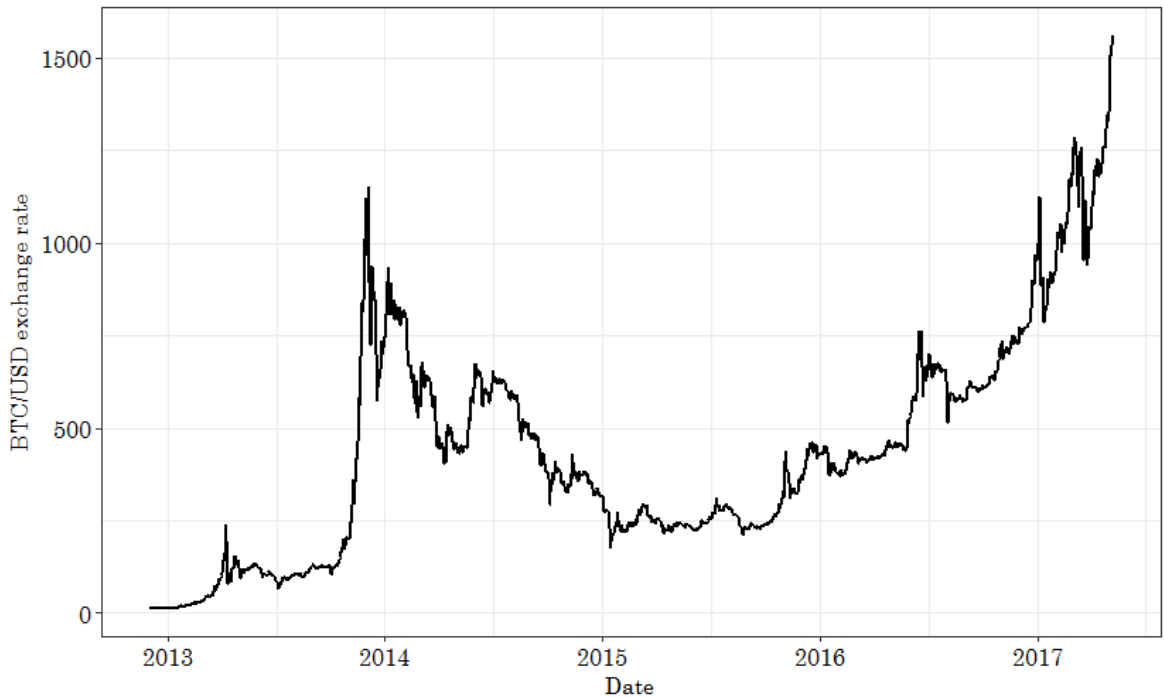

Figure 2: Bitcoin exchange rate with USD (BTC price)

In regard to the variables that are associated with the supply-demand I found, the anonymous characteristic of Bitcoin imply a limitation to analyze economic activity in the network, however, I will incorporate the variable that better serve as an internal drivers.

It is important to mention that although www.blockchain.info is a reliable source of information, for this research I found that www.quandl.com platform provides a straightforward way to extract the information from the first site, since there is an API wrapper package for R software that offers a direct interaction to this website.

Blockchain.info distinguish between three types of platform descriptors: currency statistics, block details, mining information, network activity and blockchain wallet activity, all the variables within such categories have a daily frequency. From currency statistics explanatory variables I will include the USD exchange trade volume (*trvou*) that represents the total USD value trading volume on major bitcoin exchanges. Among the block details, confirmation time (*atrct*) that records the median value that a transaction needs in order to be accepted into a block and added to the public ledger.

Regarding the mining information, I have included the hash rate (*hrate*) that measures the power of miner's machines. Finally, in order to analyze the network activity, firstly, I will consider the number of transactions per day which account for unique trades per day excluding the 100 most popular addresses.

Attractiveness' proxy in most of the papers is represented as search trends and Wikipedia articles' views, however Kristoufek (2015) found that both sources provide analogous results. This variable consists on weekly search queries for "bitcoin" word collected from Google Trends in the period January 2013 to May 2017 for 27 different countries. By providing proper filters as needed, this tool shows how regularly a particular search term is requested in comparison with the total search volume across countries and periods. The resulting number are expressed in a scaled range between 0 and 100 on a topic's proportion of all searches on all topics. The reason I decided to include several countries for search trends in opposition with other similar studies (Yelowitz & Wilson 2015; Kristoufek 2015; Bouoiyour & Selmi 2016b) is that behavior varies significantly across the series. This hypothesis was confirmed by applying the Dynamic Time Warping[5] algorithm, which let me visualize disparities across time series. For instance, the trends in China (CN), the second country in importance into trade volume of Bitcoin differs greatly from the United States (US), however, the latter seems to be more correlated with Canada (CA) and other European nations such as Great Britain (GB), Sweden (SE), and fairly stronger with France (FR), Germany (DE) and so on (figure 3).

---

[5] Dynamic Time Warping (DTW) is a technique to find optimal alignment between time dependent sequences. This method is particularly useful to measure similarity and, by extension in classification problems. For a broader explanation and application of this method please review Vaughan & Gabrys (2016) and Kate (2016).

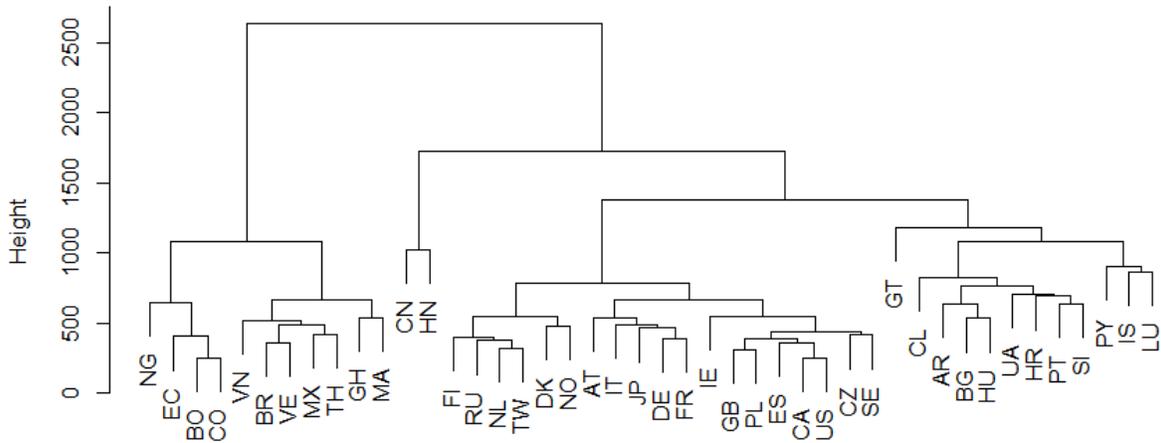

Figure 3: Clusters of attractiveness by country

Finally, the financial variables will try to capture Bitcoin's capabilities as safe-haven, diversified or hedge assets. Hence, the S&P500 (indicator of the performance of a group of relevant stock market companies), the Chicago Board Options Exchange (CBOE) Volatility Index (VIX) that expresses market's expectation of one month ahead volatility, bearish sentiment from the AAII Investor Sentiment Survey, and gold's price will be employed as a potential drivers from the financial market perspective. From the macroeconomics perspective, it is essential to account for movements in the exchanges rate of the euro with the dollar, and more relevantly, the dollar with the yuan, it might affect Bitcoin's price due to capital controls that China have been introducing in order to control speculation. China has a fundamental protagonist, since more than 90% of the bitcoins are traded with the yuan, and more than 70% of the mining takes places there (Smith 2017).

# V. Methodology

In this section, I present the Structural Time Series (also named state space form models) framework for make inferences about the variables that affect BTC price, together with a description of some side tools that allow bringing a better understanding of the problem.

## 1. Structural time series models

When we study a signal (time series) it is useful to visualize it as a product of aggregating different layers, hence, the process of decomposing each layer provides an attractive method to bring a direct individual interpretation to the model. A basic additive form of given series can be expressed as:

$$observed = trend + seasonal + error \qquad (1)$$

Therefore, a state space model (SSM) is equivalent to a dynamic system composed by a seasonal and trend elements and perturbed by random disturbances (Parmigiani et al. 2009). In his comprehensive book, Harvey (1990) highlights the salient properties of this framework due to its capability to reflect characteristics of the data, make diagnostic test and consistency with previous knowledge. By extension, structural time series (STS) framework provides the possibility to expand the information to explain the observed data by adding explanatory variables as a separate component. As expected, other relevant layers such as cycles and interventions can also be included in the model if needed[6]. Moreover, SSM allow the treatment of missing observations, inclusion of stochastic explanatory variables can be permitted to vary stochastically over time, no extra theory is required for forecasting since all that is needed is to project the Kalman filter forward into the future (Durbin and Kopman 2012). The idea behind SSM is to create a "superposition", that is, a modular set of equation in which each layers forms part of the observed stochastic process, this is the reason they are also named as SST (Parmigiani et al. 2009; Harvey 1990).

---

[6] SSM are remarkably flexible, by extension, every ARIMA model can be reformulated into an SSM form, see Prado and West (2010) Parmigiani, et al. (2009), Commandeur, et al. (2007), Durbin and Koopman (2012) for an extensive treatment of the equivalencies.

A Gaussian[7] SSM can be expressed in several notations, I have found the one presented by Durbin and Koopman (2012) and used as well in Scott and Varian (2013) and Brodersen et al. (2015) the most comprehensible:

| $y_t = Z_t \alpha_t + \varepsilon_t$ | $\varepsilon_t \sim N(0, H_t)$ | (2) |
| $a_{t+1} = T_t \alpha_t + R_t \eta_t$ | $\eta_t \sim N(0, Q_t)$ | (3) |

Then, the equation 3 is the *observation equation* where $y_t$ is a $p \times 1$ vector of observations, $Z_t$ is a known $p \times m$ matrix, $\varepsilon_t$ is an independent Gaussian random error with mean zero and variance $H_t$, and $\alpha_t$ is an unobserved $m \times 1$ vector named *state vector*. On the other side, in equation 4 is called the *state equation* which is an autoregressive model of $\alpha_t$, defined as an unobservable Markovian process which can be imprecisely measured by $y_t$. In this equation $T_t$ is a known $p \times p$ matrix called the *state/transition matrix*, $R_t$ is a $p \times m$ error control matrix (indicates which rows of the state equation have nonzero disturbance terms), and $\eta_t$ is innovation, another independent Gaussian random error with zero mean and variance $Q_t$. In summary, the idea behind SSM is that the underlying stochastic process is determined by $\alpha_t$, nevertheless, since this latent system is not observable, we have to rely on the vector of observations to solve the system. There is an initial (prior) information assumed to be known for $\alpha$ that follows a normal distribution with mean $\alpha_0$ and variance $P_0$ which is also independent of $\varepsilon_t$ and $\eta_t$ for $t = 1, 2, \ldots, n$.

Posterior inference and forecasting

For a given SSM the key task is to generate predict future observations in the unobserved states, these values are computed from conditional distributions from sequential information as it is available. In this regard, the filtering [8]process compute conditional densities $\pi(\alpha_t|y_{1:t})$ as the data arrives, that is, it estimate the current value in the state vector given the disposable information in the observation vector and generates $\pi(\alpha_{t+1}|y_{1:t+1}), \pi(\alpha_{t+2}|y_{1:t+2}), \ldots \pi(\alpha_{t+n}|y_{1:t+n})$. In this case, since we are interested in predicting BTC price, one-step-ahead (OSA) predictions of the $btcprice_{t+1}$ are based on previous data, it is needed to estimate $\alpha_{t+1}$ then, based in this value generate the observation $btcprice_{t+1}$. Finally, from a starting point $\alpha_0 \sim \pi(\alpha_0)$ it is possible to recursively compute $t = 1,2,\ldots,n$ until obtain the OSA state predictive density $\pi(\alpha_{t+1}|y_{1:t})$ and OSA predictive density $\pi(y_{t+1}|y_{1:t})$. The aforementioned problem can be solved elegantly through the Kalman filter taking advantage of the Markovian structure of the SSM and the assumption that the random state and observation vectors follow a normal distribution, as well as the marginal and conditional distributions.

---

[7] The linear SSM specify that the parameters follow a Gaussian distribution. This assumption is very sensible, however, it has computational properties that provide a simplicity for estimation.

[8] It differs from the smoothing since this problem computes recursively the conditional distribution of $\alpha_{1:t}$ given $y_{1:t}$.

One common problem that arises in SSM formulation is that system matrices $(Z_t, T_t, H_t, Q_t)$ most of the time are unknown. When all the system matrices are known it straightforward to compute densities by using Maximum Likelihood Estimators (MLE), however it gets promptly complicated when uncertainty about an unknown parameter is included (Parmigiani et al. 2009; West 1996). In this regard, a Bayesian approach provides a solution, as a consequence, simulation based methods have been gaining attention due to maximum likelihood limitations[9], and by extension one of the most prominent factor that has led to the increasing interest number of applications STS methods. For instance, Markov chain Monte Carlo (MCMC[10]) methods provide a straightforward hence powerful way to simulate posterior densities when direct methods are not available (West 1996). In particular Gibbs sampling algorithm iteratively simulate and approximate filtering densities and probabilities $\pi$, from the full conditional distributions $\pi(\alpha_{0:t}|\psi, y_{1:t})$ and $\pi(\psi|\alpha_{0:t}, y_{1:t})$ where $\psi$ is the unknown parameter. Hence, *"this approach solves at the same time the filtering, smoothing, and forecasting problems for a DLM with unknown parameters."* (Parmigiani et al. 2009).

---

[9] See (Parmigiani et al. 2009; West 1996) for a further discussion.

[10] MCMC samplers must be checked in order to prove the distributional assumptions about the simulation, and it has to be stable over several draws. In most cases reduce by thinning (eliminate the burn-in period) helps to analyze the efficiency. (Parmigiani et al. 2009; Durbin & Koopman 2012)

In general terms, the proposed framework is a powerful tool to recursively generate estimations of the problem in interest. Harvey (1990) emphasized in the parallelism of SSM with econometrics in the context of simultaneous equation systems. However, the main difference is that the restrictions imposed on it by economic theory altered the reduced form in the later, while on the former, the restrictions come not from economic theory. Instead, it is related to the desire to ensure that the forecasts reflect features of the data. In this study, I aim to introduce a improve to SSM formulation proposed in Scott and Varian (2013) and Brodersen et al. (2015) that led to select the best model out a set of possible explanatory variables. The relevance of the implementation in this study arise from the variable selection problem of the set of attractiveness drivers, this, it is expected that it will handle uncertainty about which country truly play a substantial role.

2. Bayesian variable selection

One of the most crucial aspects of SSM is the definition of the most appropriate model, hence, the inclusion of apparently clustered multiple country attractiveness indexes demands a variable selection approach that assesses for the best model variable's subset.

Variable selection has been a common problem in statistics but not too much in econometrics. During the last years econometric models commonly have a few set of "true" theory specified explanatory variables, however, nowadays the "empirical revolution" in economics has changed the vision of how to do research. One problem that stems frequently in statistics is the selection of a subset of variables in a given model for the sake of interpretability or reducing variance. Among the discrete versions we have the forward/backward stepwise selection that filters through all possible subsets, nevertheless is computationally costly when the number of predictors become large. On the other side, penalized shrinkage methods such as LASSO (Tibshirani 1996) or Ridge (Hoerl & Kennard 1970) are more generally recommended, especially in high-dimensional settings. Nonetheless, in SSM framework there is an attractive implementation that works well with MCMC and recursive estimations of the Kalman filter.

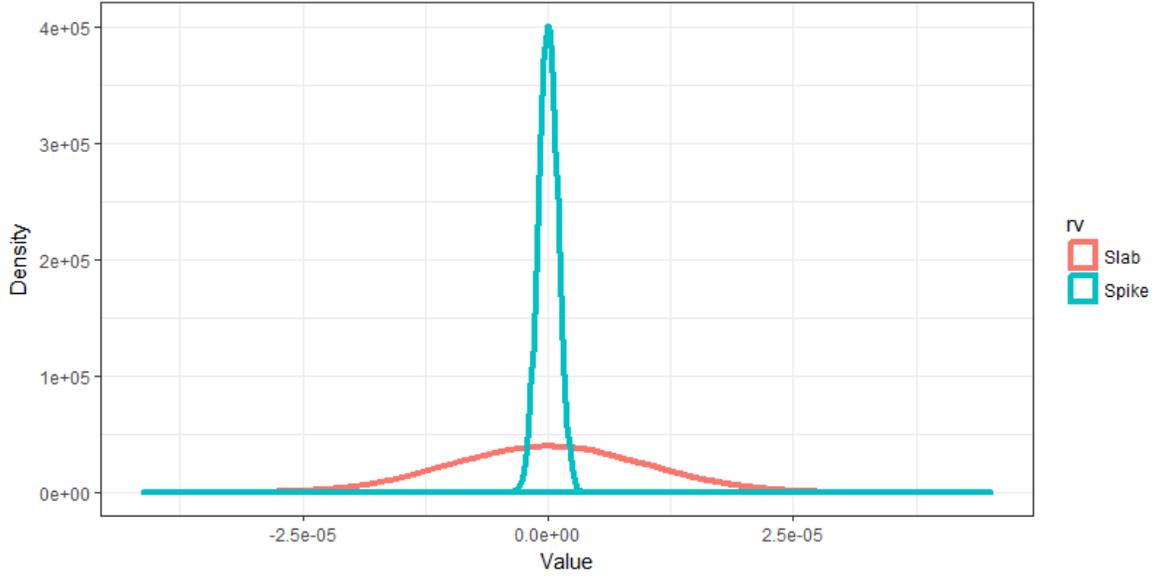

Figure 4: Spike and Slab priors

The Spike and Slab (S&S) is a hierarchical Bayesian model, where the spike refers a center of mass concentrated around or nearly to zero, while the slab is represented as a wide (high variance) normally distributed prior. As Ishwaran & Rao (2005) mentioned, these names were originally proposed by Mitchell & Beauchamp (1988), who also designed the application to follow a Gaussian distribution with the purpose of enabling efficient Gibbs sampling of the posterior conditional distributions. One popular version of the S&L model was introduced by George & McCulloch (1993) that describes the system as a mixture of two normal distributions (similar to figure 4):

$$p(\beta_j, \gamma_j) = p(\gamma_j)p(\beta_j|\gamma_j) = p(\gamma_j)\{(1 - \gamma_j)N(0, \tau_j^2) + \gamma_j N(0, c_j \tau_j^2)\} \qquad (4)$$

The idea behind S&L is to zero out $\beta_j$ coefficients that are truly 0 by making their posterior mean values small. Small hypervariances $\tau_j^2$ sets $\gamma_j = 0$ and asymptotically set the $\beta_j$ as 0, while large values of $\tau_j^2$ and $c_j$ derive into $\gamma_j = 1$, generating a non-zero estimate of $\beta_j$ which means that are going to be selected as being part of the final model. In summary, through Bayes' rule, the probabilities are updated in order to generate a joint posterior distribution of the variables with the higher marginal posterior inclusion probabilities (Scott & Varian 2013; Owusu et al. 2016; Harvey 1990). Following Brodersen *et al.* (2015).

Spike and slab priors' specification

Bayesian analysis requires explicit specification of a prior on the parameters. Non-informative priors are commonly used by researchers because it is difficult to find a universally justifiable subjective prior. However, as (Ishwaran & Rao 2005) argues, the choice of priors is often complex, although empirical Bayes approaches can be used to deal with this issue (Chipman et al. 2001). Similarly, Koop et al. (2007) supports the idea of using empirical Bayes methods to select hyper parameters values in opposition to no informative ones, nonetheless, it warns about the "double-counting" problem, that is, using the same data in previous draws to generate priors in posterior simulations. In this work it has been decided to follow an empirical selection hyper parameters selection. Hence, in the first step it will be run multiple draws of the model using the what George & McCulloch (1993) of assigning for each independent $\gamma_j$ Bernoulli($\gamma_j$) random variables a inclusion/no-zero prior probability equal to 0.5. The decision of setting such prior derives from the assumption of having no information about the presence of the variables considered, or in other words a complete indifference or uninformative priors. The second step is to generate a joint distribution from the $n$ simulations draws generated with the same model specification and inclusion prior. Finally, it is going to generate new priors and test for the convergence and sensibility.

3. Model specification

As I describe in the last chapter, the additive structure of the SSM models allow organizing the components following the idiosyncratic characteristics of the phenomena in the study. For the sake of simplicity, initially, I will present the measurement equation regular form rather than state space form[11]:

---

[11] In the appendix 3 I have describe the specification in state space form.

$$Y_t = \mu_t + \sum_{i=1}^{k} \lambda_{it}\varpi_{it} + \sum_{j=1}^{k} \beta_{jt}x_{jt} + \gamma_t + \epsilon_t \quad \epsilon_t \sim N(0, \sigma_\epsilon^2) \quad (5)$$

$$\mu_{t+1} = \mu_t + v_t + \xi_t \quad \xi_t \sim N(0, \sigma_\xi^2) \quad (6)$$

$$v_{t+1} = v_t + \zeta_t \quad \zeta_t \sim N(0, \sigma_\zeta^2) \quad (7)$$

$$\lambda_{i,t+1} = \lambda_{i,t} + \rho_{i,t} \quad \rho_{i,t} \sim N(0, \sigma_\rho^2) \quad (8)$$

$$\beta_{i,t+1} = \beta_{i,t} + v_{i,t} \quad \tau_{i,t} \sim N(0, \sigma_v^2) \quad (9)$$

$$\iota_{1,t+1} = -\iota_{1,t} - \iota_{2,t} - \iota_{3,t} + \varsigma_t \quad \varsigma_t \sim N(0, \sigma_\varsigma^2) \quad (10)$$

$$\iota_{2,t+1} = \iota_{1,t} \quad (11)$$

$$\iota_{2,t+1} = \iota_{2,t} \quad (12)$$

Regarding the variances $\sigma^2$, they are typically modelled as an inverse gamma distribution of the precision $(1/\sigma^2)$, hence:

$$\frac{1}{\sigma^2} \sim Gamma\left(\frac{v}{2}, \frac{s}{2}\right) \quad (13)$$

Where $\mu_t$ is the local level component, this component is analogue to the intercept in a classical regression model with the different of being able to change over time, while $v_t$ represents the angle of the trend line that also varies over time. The $i$ intervention or shocks variables that are going to be included in the model are denoted as $\lambda_{it}$ for $i = 1, \dots, k$, this component will capture suddenly changes in the level at the time point where the event happened. There are three possible situations after an intervention, a level shift which means a permanent structural form in the series, slope shift which means that the value of the slope showed a permanent change after the intervention and finally pulse, where the value of the level suddenly changes at the moment of the intervention, and immediately returned to the value before the intervention. In order to study the effects of other variables, a set of explanatory variables are going to be included where $\beta_{jt}$ is an unknown regression weight for $j = 1, \dots, k$. Finally, $\iota_t$ captures the seasonal component of the series, in this case it is expected that BTC price has quarterly periodicity.

4. Standardized variables

Standardization is the process of taking the sample mean of a random variable and dividing the result by its standard deviation.

$$\frac{Y_i - \overline{Y}}{S_Y} = \left(B_1 \frac{S_1}{S_Y}\right) \frac{X_{i1} - \overline{X}_1}{S_1} + \cdots + \left(B_k \frac{S_k}{S_Y}\right) \frac{X_{ik} - \overline{X}_k}{S_k} + \frac{E_i}{S_Y} \quad (14)$$

The use of standardization of the covariates and response variable has been part of a long time discussion in statistics. Detractors' main critic is around the use of standardized coefficients as a comparative importance measure among a different class of variables due to the "unitless" property of standardized variables. Moreover, when independent and dependent variables differ greatly from their distribution. (Nimon & Oswald 2013; Greenland et al. 1991). However, it offers a set of advantages. Firstly, in a regression model, a coefficient of unstandardized variables measures the expected change in the dependent variable when the independent variable change in one unit. Conversely, when both variables are standardized, the interpretation differs, thus, the modified coefficient measures the expected standard deviation variation in the response variable associated with one standard deviation change in the covariate. In time series analysis studying the variance movements makes more sense than levels, hence standardized coefficient offer and attractive characteristic, beyond the comparability across different type of variables that in this research presents. Secondly and more importantly, in this study we consider predictors with a considerable level of collinearity (search trends) "*which intercorrelations between predictors (multicollinearity) undermine the interpretation of MLR weights in terms of predictor contributions to the criterion*" (Nimon & Oswald 2013) and standardized variables help model selection in the presence of S&L approach since as it was stated reduces the variability of the estimates by shrinking the coefficients and reduction of collinearity (Oyeyemi et al. 2015; Clyde 1999; Ročková & George 2014)

5. Assessing seasonality

Several authors have concluded that Bitcoin time series seems to behavior unlike any other asset, this characteristic demands a closer look at the structure of trend and seasonal components. In order to test for the latter component, I will provide a periodogram that is used to identify the dominant periods (or frequencies) of a time series. This might be a helpful tool for identifying the dominant cyclical behavior in a series, particularly when the cycles are not related to the commonly encountered monthly or quarterly seasonality (Shumway & Stoffer 2010).

# VI. Forecasting results

In this section, it is going to be compared to one-step-ahead forecasts estimates with structural time series method and actual series for the period 01/2013 until 05/2017 following the superposition of the components. It has been started with a naïve local level model without explanatory variables (the basic form in the state space framework), where the unobserved level $\mu_t$ (equation 7) has an irregular component hence, defined as random walk with the form $\mu_{t+1} = \mu_t + \xi_t$. As it can be noticeable in the local level model the equation 7 assumed that $v_t$ is zero, this term is now included as a new state equation for modelling the slope (also called drift) $v_t$ which measures the angle of the stochastic trend line. Regarding seasonal component it has been discovered that BTC price does not have a recurring pattern over time, this conclusion arises from the periodogram analysis (figure 10) where the highest periodic signal peak appears at 960, which is in this case meaningless given that the time series number of observations are 1620. The now-casting performance will be described in table 1 by different typical accuracy measures.

Table 1: In-sample prediction accuracy according to different specification

| Model | | sMAPE | MAE | MSE |
|---|---|---|---|---|
| Local level | LL | 3.146 | 12.749 | 506.992 |
| Local level with time-invariant regressors | LLTI | 4.874 | 12.139 | 457.650 |
| Local level with time-variant regressors | LLTV | 4.181 | 14.588 | 702.588 |
| Local linear trend | LLT | 2.970 | 12.026 | 499.041 |
| Local linear trend with time-invariant regressors | LTTI | 4.134 | 11.782 | 455.146 |
| Local linear trend with time-variant regressors | LLTTV | 3.825 | 12.730 | 540.861 |

The process of superposing features in a structural model provides a flexible, tractable and intuitive process to analysis the behavior of BTC price. Prediction accuracy results for the period 01/2013 to 05/2017 are presented and compared in the table above. According to the symmetric mean absolute percentage error (sMAPE[12]), the naïve local linear trend model provides the best fit to the in-sample data with an error of 2.970%, followed by the local level of 3.146%. The effects of introducing regressors decrease

---

[12] The symmetric mean absolute percentage error (sMAPE) is an accuracy measure based on relative errors, it is evaluated as: $\frac{100}{n}\sum_{t=1}^{n}\frac{|\hat{Y}_t - Y_t|}{(|Y_t| - |\hat{Y}_t|)/2}$

precision when we account only for sMAPE as an indicator, however, predictors let to analyze the association with potential drivers, which is one of the objectives of this study. On the other side, the mean absolute error (MAE[13]) and the mean squared error (MSE[14]) indicators, points to the LLTTI model since it has the lowest value for both, followed by LLT model. In this study, it has been decided to follow the superposition of LLTTV model, given that it provides the prospect to dynamically analyze the association of price drivers.

Figure 5: One step ahead predictions of Bitcoin's price

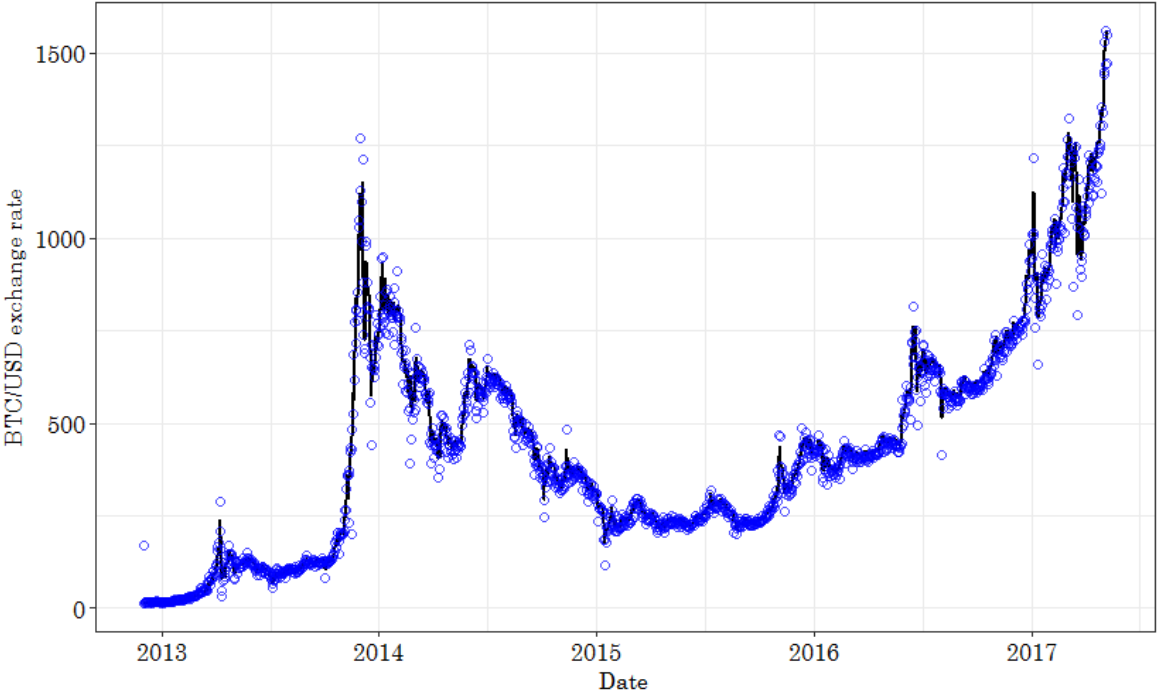

Note: Blue dots describe the one-step-ahead predictions for Bitcoin's price.

Prediction results for the LLTTV model are shown above. Here it is seen that the model predicts reasonably well, however when it tends to overestimate local maxima and local minima values, this is one of the reasons why the MSE measure was relatively bad in comparison with other specifications since the square weights heavily the presence of extreme values. This aspect is confirmed by the SMAPE measure in Table 1, that locate in terms of prediction power on the third position.

---

[13] The mean absolute error (MAE) is as its name describe $\frac{\sum_{t=1}^{n}|e_i|}{n}$

[14] The mean squared error (MSE) is represented as $\frac{\sum_{t=1}^{n}(e_i)^2}{n}$

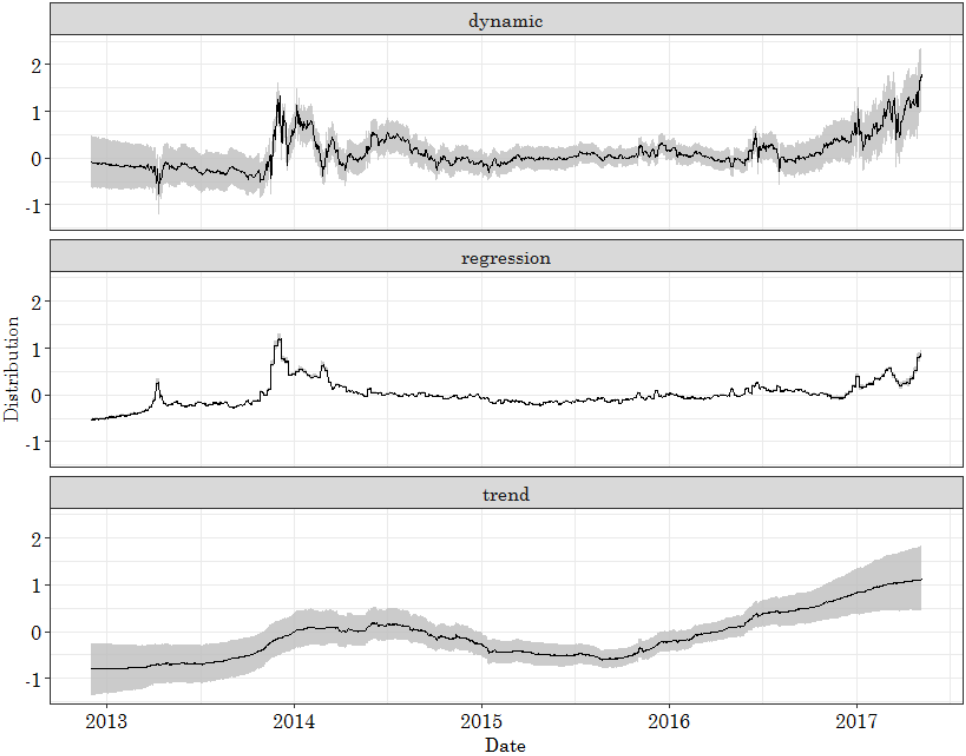

Figure 6: Decomposition of the LLTTV model for Bitcoin price

Note: the gray ribbon shows the 95% HDI.

As it has been mentioned, one of the main features of the state space framework is its ability to decompose the forecast into diverse components. The figure 5 above illustrates the contribution of each of those components, where we can highlight the slope trend signal estimated recursively by the Kalman filter. The medium panel shows the effect of time-invariant regressors that provided prediction power to explain Bitcoin price by the end of 2013 and beginning of 2014, with an overall level of uncertainty (perceived by vaguely noticeable gray ribbon). Similarly, the time-variant regressors contributed heavily to price variation, mainly in the first semester of 2014 and 2017 up to the end of the period of analysis.

# VII. Posterior estimates results

We have conducted an estimation of Bitcoin's price based on a set of internal and external demand factors. However, it is precise to organize the empirical analysis in five sections in order to present clearly the process of the methodology. In the first part I will provide a description of the hyperparameters and prior calibration, in the second part, I describe the results of the variable selection procedure, thirdly, the static and dynamic coefficient estimates and final the prediction comparison across models.

### 1. Hyperparameters and priors calibration

In the variable selection section, it was stated that the Spike and Slab method discriminate coefficients based on the values of $\tau$, where the prior for this parameter follows a Bernoulli distribution. In this study it has been run 30 different MCMC simulations with 3000 (discarded 981 draws as they represent burn-in period) iteration each (graph 2, appendix) with an uninformative prior of 0.5 as the authors suggested, and a prior mean equal to 0 for all variables. This process was done in computationally terms it is imperative to use multiple "seeds", that is, different random number generators in order to learn about the "true" posterior probabilities.

In the graph 2a it can be denoted that there are a group of variables that commonly have a fairly stable non-zero coefficient, that is the case of gold price, S&P500 and search trends from Colombia (trend_co), while Hash Rate of VIX is asymptotical to zero and unstable. Since one of the most attractive characteristics of the Spike and Slab approach is the possibility to learn from the posterior distribution, and incorporate the information as a prior for further analysis, the best guess for the prior to incorporate in a unique MCMC simulation used as a base for the model is to mean inclusion probability a coefficient means values. This process is usually called as "empirical Bayes" that is, utilize previous information results to "shrinkage" subsequent simulations[15]. This procedure intends to address the problem of limited computational power to estimate through a single MCMC "true" posterior probabilities, hence, I integrate information at multiple simulations to provide more accurate inference for the reference model[16].

---

[15] Stability of the further simulation was proved given the update of the priors (graph 3a)
[16] Extension of the process are described in detail in Xi, Li, Hu, & others (2016).

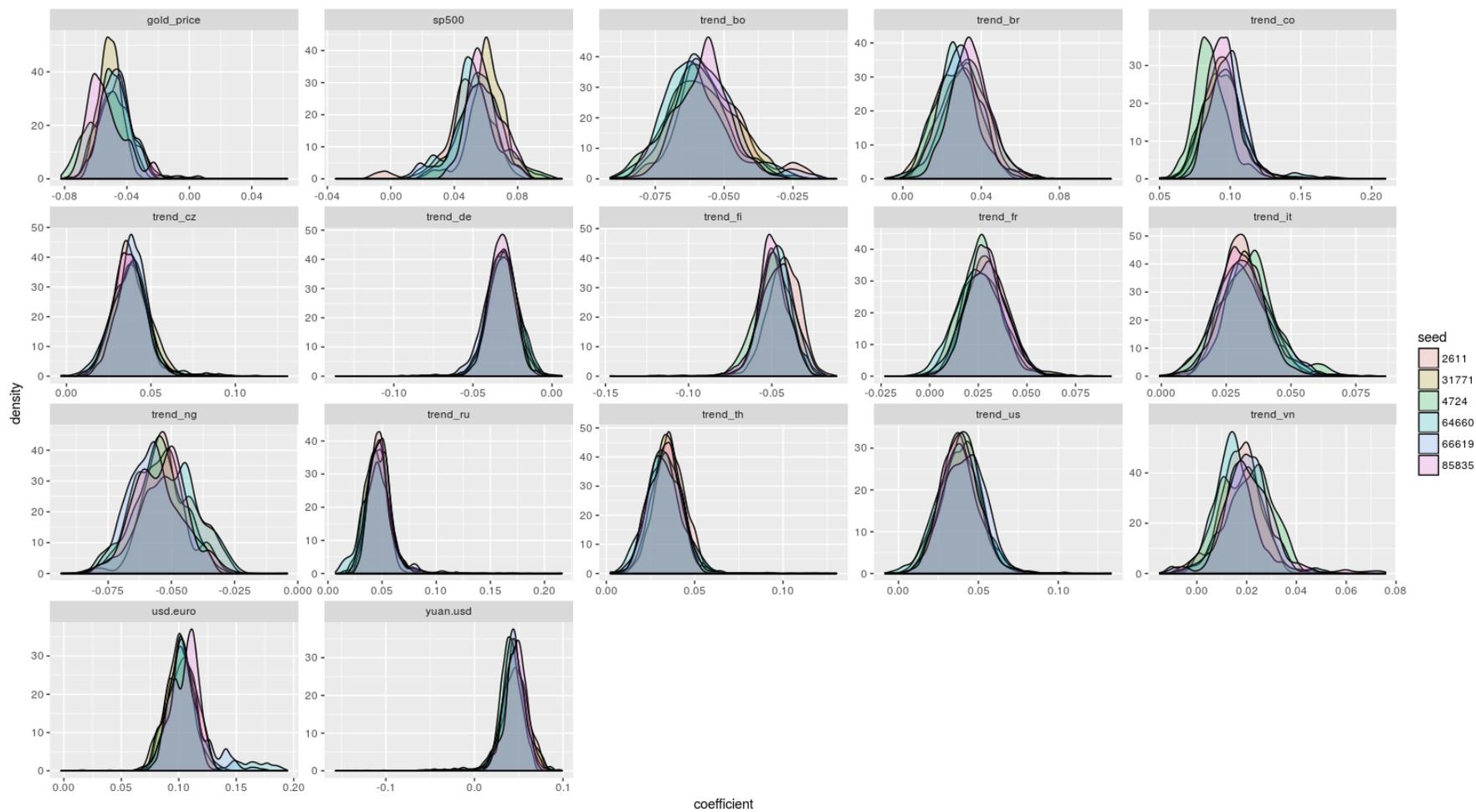

Figure 6: Multiple simulation posterior densities for the non-zero inclusion probability variables

The graph 4 shows the density of the variables whose mean inclusion probabilities surpassed the 80% for six distinct simulations with 10000 MCMC iterations (discarding a burn-in it would effectively 8129).

As it was stated before, previous research on this topic has been conducted using search trends as an attractiveness proxy, but only considering the signal of all countries on Google, however, the behavior varies significantly across countries. Hence, the variable selection procedure helps us to discern which series have higher prediction power.

2. Marginal posterior regression estimates

Following the methodology, I estimate a Bayesian Structural Times Series model of Bitcoin's price drivers, in this model the dependent variable is the standardized level of Bitcoin's price, not the returns. Table 1 summarizes the statistics associated with estimating the change in the standard deviation of Bitcoin's price given one standard deviation change in the. Estimation was constructed from the updated prior, marginal posterior means, medians (both are nearly identical since the distributions are seemly symmetrical), 95% highest density intervals (HDI) as well as non-zero probability. The first thing to note regarding the results in Table 1 is the zero in the intercept, this result derives from the standardization of the variables, additionally we can see the presence 17 relevant variables out of the 55 that we have considered in first place. Additionally, it has been estimated the dynamic coefficient for those variables that have a relevant non-zero probability of being part of the model.

Table 2: Time-invariant statistics of the standardized coefficients

| Variable | Mean | 2.5% | 97.5% | Non-zero probability |
|---|---|---|---|---|
| (Intercept) | 0.000 | 0.000 | 0.000 | 0.000 |
| Median conf. time | 0.000 | -0.002 | 0.003 | 0.007 |
| Bearish sentiment | 0.000 | -0.003 | 0.001 | 0.008 |
| Gold's price | -0.051 | -0.062 | -0.040 | 1.000 |
| Hash rate | 0.000 | -0.011 | 0.009 | 0.016 |
| My wallet trans. | 0.000 | -0.003 | 0.006 | 0.014 |
| Trans. excl. popular | 0.000 | -0.006 | 0.016 | 0.023 |
| S&P500 | 0.057 | 0.044 | 0.069 | 1.000 |
| Trend Argentina | 0.000 | -0.011 | 0.001 | 0.027 |
| Trend Austria | 0.000 | -0.011 | 0.013 | 0.019 |
| Trend Bulgaria | 0.000 | -0.013 | 0.004 | 0.026 |
| Trend Bolivia | -0.058 | -0.073 | -0.041 | 1.000 |
| Trend Brazil | 0.042 | 0.028 | 0.056 | 1.000 |
| Trend Canada | 0.000 | -0.010 | 0.013 | 0.019 |
| Trend Chile | 0.000 | -0.007 | 0.006 | 0.018 |
| Trend China | 0.000 | -0.002 | 0.001 | 0.005 |
| Trend Colombia | 0.092 | 0.075 | 0.105 | 1.000 |
| Trend Czech Republic | 0.041 | 0.027 | 0.054 | 1.000 |
| Trend Germany | -0.033 | -0.047 | -0.019 | 1.000 |
| Trend Denmark | 0.000 | -0.008 | 0.007 | 0.011 |
| Trend Ecuador | 0.000 | -0.006 | 0.005 | 0.009 |
| Trend Spain | 0.000 | -0.006 | 0.007 | 0.012 |
| Trend Finland | -0.038 | -0.051 | -0.025 | 1.000 |
| Trend France | 0.042 | 0.028 | 0.056 | 1.000 |
| Trend United Kingdom | 0.000 | -0.007 | 0.011 | 0.014 |
| Trend Ghana | 0.000 | -0.009 | 0.003 | 0.013 |
| Trend Guatemala | 0.000 | -0.002 | 0.001 | 0.003 |
| Trend Honduras | 0.000 | -0.001 | 0.002 | 0.003 |
| Trend Croatia | 0.000 | -0.007 | 0.006 | 0.015 |
| Trend Hungary | 0.000 | -0.007 | 0.005 | 0.021 |
| Trend Ireland | 0.000 | -0.007 | 0.006 | 0.010 |
| Trend Iceland | 0.000 | -0.002 | 0.002 | 0.005 |
| Trend Italy | 0.031 | 0.018 | 0.044 | 1.000 |
| Trend Japan | 0.000 | -0.007 | 0.008 | 0.011 |
| Trend Luxembourg | 0.000 | -0.002 | 0.007 | 0.016 |
| Trend Morocco | 0.000 | -0.006 | 0.008 | 0.019 |
| Trend Mexico | 0.000 | -0.005 | 0.006 | 0.007 |
| Trend Nigeria | -0.057 | -0.070 | -0.044 | 1.000 |
| Trend Netherlands | 0.000 | -0.005 | 0.013 | 0.015 |
| Trend Norway | 0.000 | -0.006 | 0.004 | 0.010 |
| Trend Poland | 0.000 | -0.010 | 0.007 | 0.011 |
| Trend Portugal | 0.000 | -0.002 | 0.006 | 0.011 |
| Trend Paraguay | 0.000 | -0.002 | 0.004 | 0.005 |
| Trend Russian Federation | 0.051 | 0.036 | 0.066 | 1.000 |
| Trend Sweden | 0.000 | -0.010 | 0.005 | 0.015 |
| Trend Slovenia | 0.000 | -0.004 | 0.002 | 0.006 |
| Trend Thailand | 0.036 | 0.024 | 0.050 | 1.000 |
| Trend Taiwan | 0.000 | -0.005 | 0.004 | 0.009 |
| Trend Ukraine | 0.000 | -0.011 | 0.004 | 0.016 |
| Trend United States | 0.054 | 0.039 | 0.069 | 1.000 |
| Trend Vietnam | 0.000 | -0.006 | 0.008 | 0.013 |
| Trend Venezuela | 0.031 | 0.020 | 0.041 | 1.000 |
| Exchange Trade Volume | 0.000 | 0.000 | 0.012 | 0.039 |
| USD-Euro exchange rate | 0.099 | 0.088 | 0.112 | 1.000 |
| VIX | 0.000 | -0.003 | 0.003 | 0.009 |
| YUAN-USD exchange rate | 0.039 | 0.027 | 0.054 | 1.000 |

Results on internal determinants

Among the internal variables that are directly related to bitcoin, I have not found any relevant effect on bitcoin's price. It was expected that the daily median time take for transactions to be accepted into a block would have a negative association with the price, however, one change in the standard deviation had zero effect, with an inclusion probability of just 0.07%. Similarly, the hash rate which measures the productivity and difficulty of the blockchain had had a very low non-zero probability of being part of the model. This result differs from Kristoufek (2015) and Georgoula et al. (2015) who found a positive, however, small (and diminishing in the case of the former) effect on bitcoin's price. Other variables such as transactions excluding popular addresses and trade volume had a slightly higher inclusion probability (2.3% and 3.9%) than hash rate and confirmation time, however negligible.

Results on attractiveness determinants

The previous part of the discussion concerns the results obtained for the internal factors set, in this section, I will provide the outcomes for the attractiveness or interest for Bitcoin. Before starting the analysis it is precise to highlight that popularity of Bitcoin, and by extension the interest measured from search trend indisputably several limitations as an attractiveness proxy, first, we do not know the true reasons why people from different countries look on the internet. Second, the fact that a person is interested in gaining information not necessary means that that is going to actively participate in the market. Nonetheless, given that it is almost impossible to locate where the transactions take place (Athey et al. 2016), search trends provide a good approximation.

As it was stated, there is a great difference in the behavior of the signal across countries, an aspect that has not been analyzed in empirical studies. Hence, this characteristic can be interesting and provide more detailed results since the different governments have been developing policies, either for providing a legal framework or to limit the use of Bitcoin. Among the countries in consideration for the initial model 13 out of 44 had a high probability of being part of the final model, additionally, the interest appears to differ significantly in sign and magnitude amid the selected series.

The marginal posterior mean and HDI for Colombia is 0.092 [0.075, 0.105], that is, 1 standard deviation change in the searches for "Bitcoin" in google from this country is associated with almost 0.1 standard deviations change in Bitcoin's price. Furthermore, the aforementioned country has a common cluster's leaf with other two selected trends which nearly identical posterior means, Bolivia (-0.058 [-0.073, -0.041]) and Nigeria (-0.057 [-0.070, 0.044]). Both countries' governments have emitted a formal warning about the use of bitcoin to make transactions since it is not a legal tender. The dynamic regression results provide a more accurate description of the marginal probability distribution over time of the standard deviation effect of search trends. For instance, the coefficient of Bolivia switched from positive to negative ending 2014, while Nigeria negative effect has been fading since 2013.

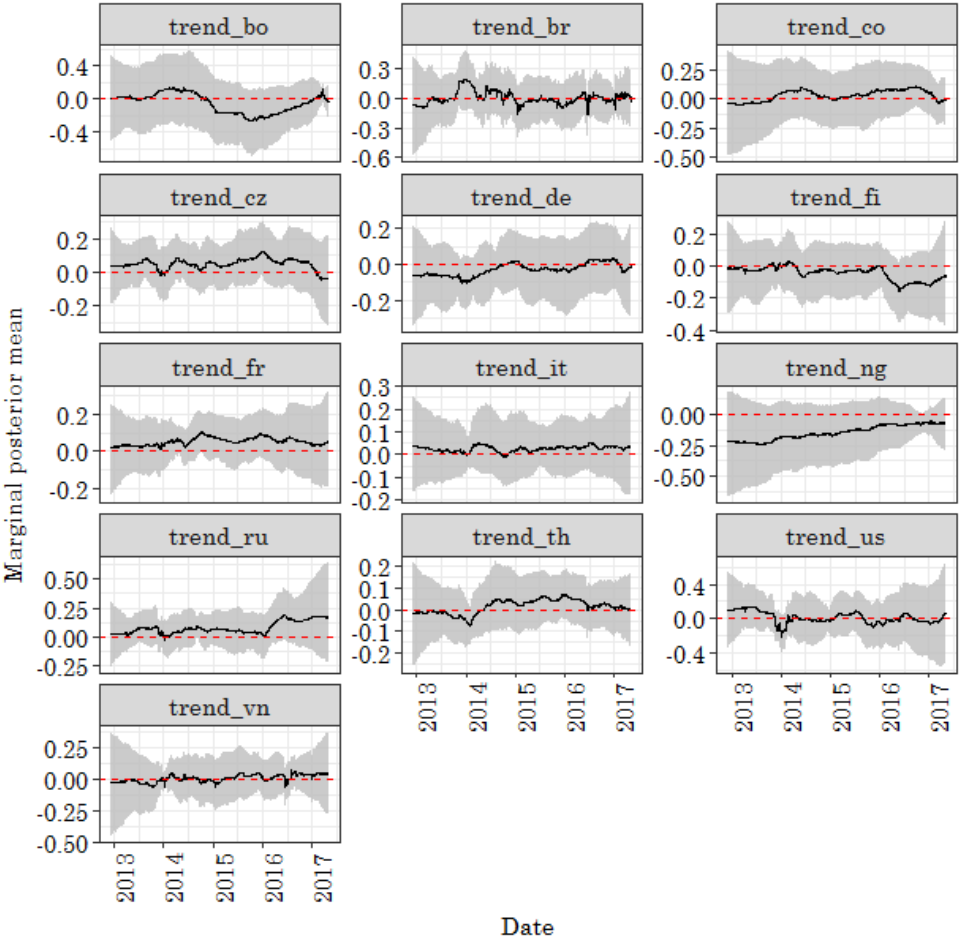

Figure 7: Dynamic standardized coefficients for attractiveness drivers

Note: Marginal posterior means (black solid line) and 95% HDI (gray ribbon).

Among the marginal posterior means estimates Russia (RU) has one the highest effects on of the price with 0.051 [0.036, 0.066] (Table 1), with a special aspect to consider.

Beginning 2016, the marginal posterior mean jumped to 1 standard deviation in the search trends from Russia has been linked with nearly 0.15 standard deviations change in Bitcoin's price. This behavior is not spurious since Russia alongside Japan has been the main two key countries in the discussion for creating or not a legal framework for the use of Bitcoin as a currency as well as other cryptocurrencies. Another country where BTC has an important relevance is Venezuela (VN), where the time-invariant marginal posterior mean shows that 1 SD change in a number of queries in Google is associated with 0.031 [0.020, 0.041] SD's in the BTC price. According to the news[17], the political and economic conjuncture, loss in government fate and uncertainty, have been producing interest in Bitcoin for its use on multiple purposes. First, "bitcoin miners" job is especially attractive since the earnings are protected from bolivar's (Venezuela's currency) extreme inflation, second, buying basic necessities (after being exchanged for dollars) inside and outside the country and as a safe haven asset. As it can be noted in figure 10, the trade volume of bitcoin has been increasing rapidly after 2016. Regarding United States (US), France (FR), Italy (IT) and Czech Republic (CZ) the results indicate a clear-cut positive association of Bitcoin's volatility, with a higher participation of US with 0.054 [0.039, 0.069] standard deviation change in the price. The time-variant marginal posterior means (figure 5) show for this group positive relationship over in the years 2013-2017 but on the other hand, it can be distinguished a short period of negative impact was evidenced in 2014, especially in the US where it reached -0.2 standard deviation change.

Results on macro financial determinants

One of the main goals of this paper is to prove if Bitcoin is a speculative asset, in such case its behavior has to be related to other assets or market indexes. Similarly, macroeconomic variables such as exchange rate also play a relevant role into uncovering the current use of BTC. By now I have proved that internal factors play are not relevant, while attractiveness in almost half of the countries does present a high probability of being part of the final model that describes standard deviations movements on BTC price. Although several attractiveness variables may have contributed to answering the question if drivers also played a significant role. Among the variables in this group, I have found that the Chicago Board Options Exchange (CBOE) Volatility Index (VIX) together with Bearish sentiment were excluded from the final model with only 0.8% and 0.9% probability of being selected (Table 1). This result differs from Bouri et al. (2017) outcomes since the authors found that uncertainty has a significant negative impact on BTC returns and by extension being a hedge against that uncertainty. However, the results of this study show that the gold price, S&P500 and bilateral exchange rates are related to Bitcoin price.

---

[17] Articles referring details about this situation can be reviewed in "The Guardian" Growing number of Venezuelans trade bolivars for bitcoins to buy necessities, "Business insider", Venezuela is cracking down on 'bitcoin fever' and "The Washington Post" Bitcoin 'mining' is big business in Venezuela, but the government wants to shut it down.

Even though Bitcoin is a recent invention, it has been gaining attention as an investment asset. Correspondingly, Bouoiyour and Selmi (2016a) theorize that even though there is plenty literature of how precious metals can act as a safe haven or hedge, few authors have tried to answer if Bitcoin behaves such as gold or silver under risk situations in the stock market given their fair stability over time. This hypothesis is shared by Bouri et al. (2017) and Dyhrberg (2016) who states that "*generally economists have compared bitcoin to gold as they have many similarities*". Additionally, Bouoiyour and Selmi (2016a) argue that transcendental political events (Trump election specifically) might generate stock market due to geopolitical uncertainty, an event that could generate interest in Bitcoin.

According to the results shown in Table 1, one SD change in the gold's price is linked to -0.051 [-0.062, -0.040] SD's change in BTC price, with an inclusion probability of 100%. Henceforth, given that a hedge is an asset that is marginally negative correlated with another asset (Baur & Lucey 2010), the coefficient suggest that Bitcoin acts as a hedge in relation to the gold, plus the prospect characteristic of being the "digital gold", this result are similar to Dyhrberg (2016). The results in figure 6 indicate that the dynamic coefficient has been typically negative during the period in the study, where the effect started to diminish from the third quarter of 2014 and regaining its previous state until the first quarter of 2017, with in between can be noticeable a positive standard deviation variation opening 2016.

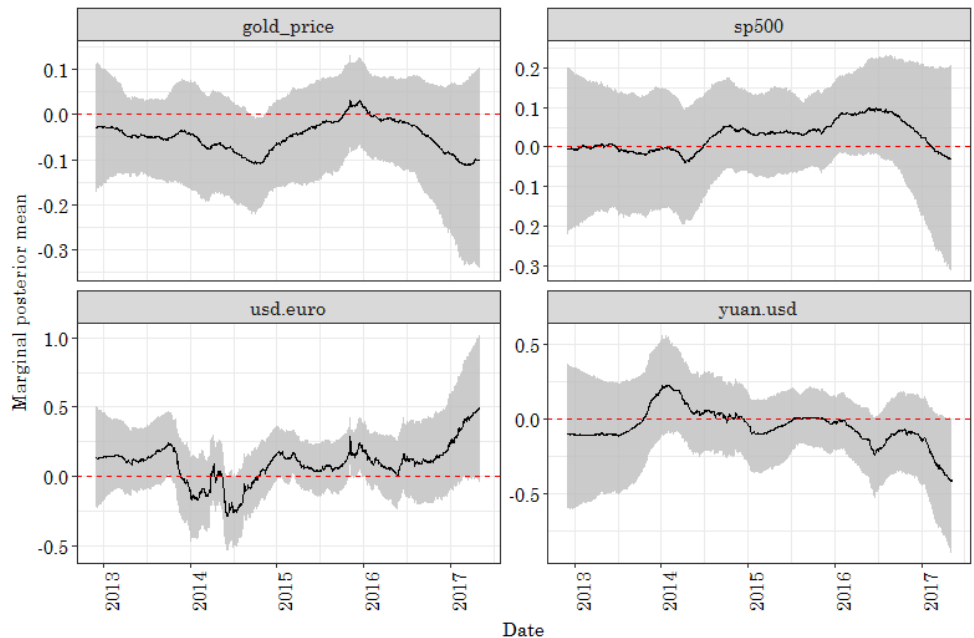

Figure 8: Time-variant standardized coefficients for macro financial drivers

Note: Marginal posterior means is described for the black solid line and 95% HDI by the gray ribbon.

Looking at the relationship to the stock market indicator, the model shows that one SD variation in the S&P500 index is on average associated with 0.057 [0.044, 0.069] SD change in BTC price. However, from the time-variant estimation (figure 6) it can be noticeable the negative association (yet small) from 2013 until the second semester of 2014, posteriorly there is a switch in the sign that lasted up to 2017, then it got negate again. From the jargon proposed by Baur and Lucey (2010), it can be determined that Bitcoin has been performing as a diversifier and hedge, with a tendency converge at the end of the period of analysis to the latter.

Regarding exchange rates, there are two aspects that deserve further attention, first, following credibility intervals it is straightforward to conclude that the effect on Bitcoin is reasonably stronger in comparison with any other price driver studied. Second, there is a substantial difference both in the sign and behavior of time-variant coefficients over time of both bilateral exchange rates into consideration. The results suggest that one SD change in the USD-Euro exchange rate is linked to 0.099 SD's change in BTC price, with a 95% probability that this value is going to be positioned in the range 0.088-0.112. On the other side, one SD change in the Yuan-USD exchange rate is associated with 0.039 [0.027, 0.054] change in the SD of BTC price. Nonetheless, dynamic coefficients provide a more informative perspective of BTC relationship with exchange rates, which is the case for the last two months of 2013 when the price reached $1242 per BTC (surpassing for a small period of time gold´s price), incident that attracted global attention towards cryptocurrencies. The lowest row of graphs in figure 6 depict the effect of the sudden rise in BTC price on exchange rates, in USD-Euro case, the effect switched signs from nearly 0.23 to -0.17 SDs in BTC price given one SD change in USD-Euro exchange rate, followed BTC price behavior in the second rise that took place in midst 2014. Regarding Yuan-USD exchange rate[18] one SD increase in this variable leads to an increase of 0.039 [0.027, 0.054] SDs in BTC price. However, this might seems deceptive given results exposed on figure 6, it happens due to the dependency of the prior selection, and hence the time-variant marginal posterior provides a proper interpretation of the relationship. As it can be described in the graph, a positive unit change in the SD in Yuan-USD leads over most of the time to a negative impact on BTC price volatility, with an acceleration by the end of the period of analysis.

From a macro-financial factor driver's point of view, it seems that BTC price variation is more sensible to exchange rates than stocks market deviations. Additionally, the results seem to be related to Dyhrberg (2016) in the sense that BTC has properties that range between a currency and a commodity.

---

[18] This value can be misleading since as depicted in figure 9 in the appendix, the marginal posterior distribution for Yuan-USD is bimodal, that is, with two local maxima, one negative around -0.03 and other positive around 0.08. Since the Slab component on S&L variable selection is normally distributed, I have decided to keep the prior distribution as is, and set a naïve approximation for the mean of the bimodal distribution.

# VIII. Conclusion

I have presented the results of now-casting for bitcoin exchange rate with the dollar, additionally, it has been studied price drivers. In general, the research revealed that Bitcoin possesses a multifaceted property that goes between a currency, hedge and safe haven assets for geopolitical instability, and presumably a payment method.

The dynamic analysis of the BTC price has yielded new insights about the relationship between different collections of factors. It has been speculated that Bitcoin might be entering in a new phase, in this regard the increasing effect of attractiveness may be indicative prospect of such argument, and also the consequences of signals from government's policies to find a legal framework. The results expressed here support both of these hypotheses, as we found evidence for interest in critical countries.

[ADD]

http://www.tandfonline.com/doi/abs/10.1080/13504851.2014.995359.

Yermack, D., 2013. *Is Bitcoin a Real Currency? An Economic Appraisal*,

# X. Appendix

1. Appendix 1: Gibbs sampling

## 2.4. Markov Chain Monte Carlo Algorithm

Posterior inference and function selection is based on a blockwise Metropolis-within-Gibbs sampler. The sampler cyclically updates the nodes in Figure 1. For Gaussian responses it reduces to a Gibbs sampler. The full conditionals of the parameters $w$, $\tau_j^2$, $\gamma_j$, $j = 1, \ldots, p$, and the means $m = (m_1, \ldots, m_l, \ldots, m_q)'$ of the conditionally Gaussian variables $\xi | m_l \sim N(m_l, 1)$, $m_l = \pm 1$, are available in closed form and are included in Algorithm 1 in the appendix. They do not depend on the specific exponential family chosen for the responses.

The full conditionals for $\alpha = (\alpha_1, \ldots, \alpha_p)'$ and $\xi = (\xi_1', \ldots, \xi_p')'$ depend on the "conditional" design matrices $X_\alpha = X \operatorname{blockdiag}(\xi_1, \ldots, \xi_p)$ and $X_\xi = X \operatorname{diag}(\operatorname{blockdiag}(1_{d_1}, \ldots, 1_{d_p})\alpha)$, respectively, where $1_d$ is a $d \times 1$ vector of ones and $X = (X_1, \ldots, X_p)$ is the concatenated design matrix. For Gaussian responses, the full conditionals are given by

$$\alpha | \cdot \sim N(\mu_\alpha, \Sigma_\alpha) \text{ with}$$

$$\Sigma_\alpha = \left(\frac{1}{\sigma^2} X_\alpha^T X_\alpha + \operatorname{diag}(\gamma \tau^2)^{-1}\right)^{-1}, \quad \mu_j = \frac{1}{\sigma^2} \Sigma_\alpha X_\alpha^T y, \text{ and} \tag{5}$$

$$\xi | \cdot \sim N(\mu_\xi, \Sigma_\xi) \text{ with}$$

$$\Sigma_\xi = \left(\frac{1}{\sigma^2} X_\xi^T X_\xi + I\right)^{-1} ; \quad \mu_j = \Sigma_\xi \left(\frac{1}{\sigma^2} X_\xi^T y + m\right).$$

2. Appendix 2: Graphics

Figure 9: Periodogram of Bitcoin's price

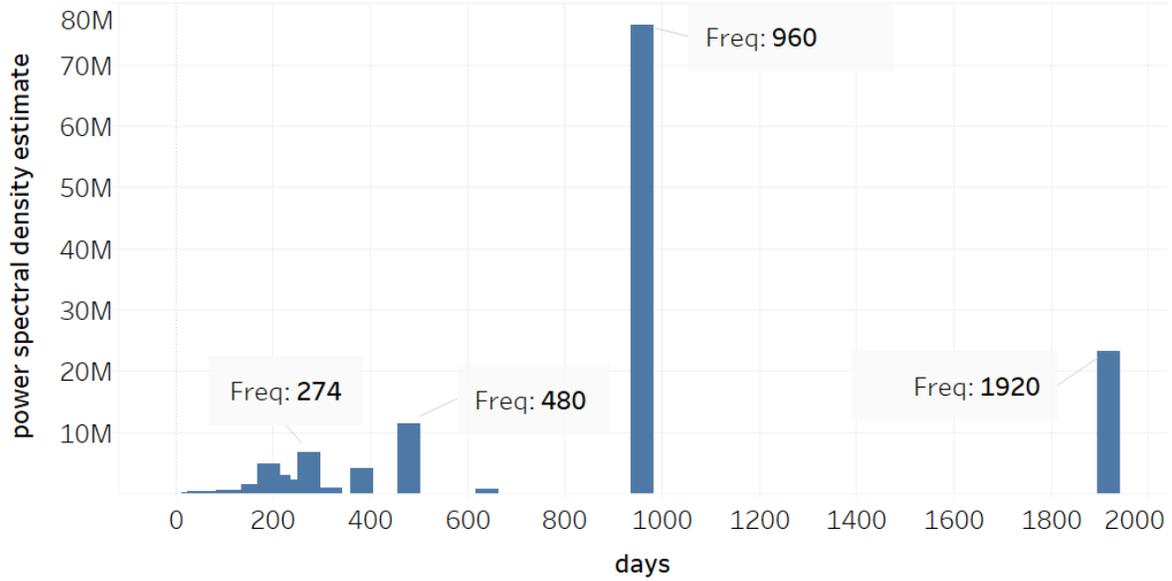

Figure 10: Mean search trends values by cluster

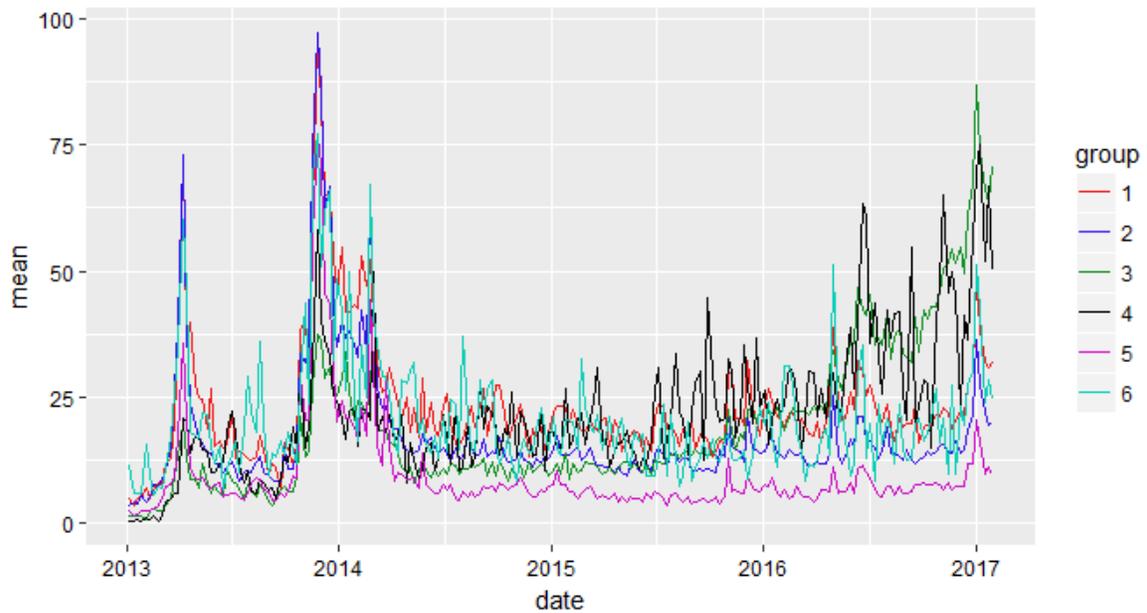

Figure 11: Markov chain and Monte Carlo Simulations for time-invariant $\beta$ coefficients

Figure 12: Marginal posterior distributions employed to generate priors

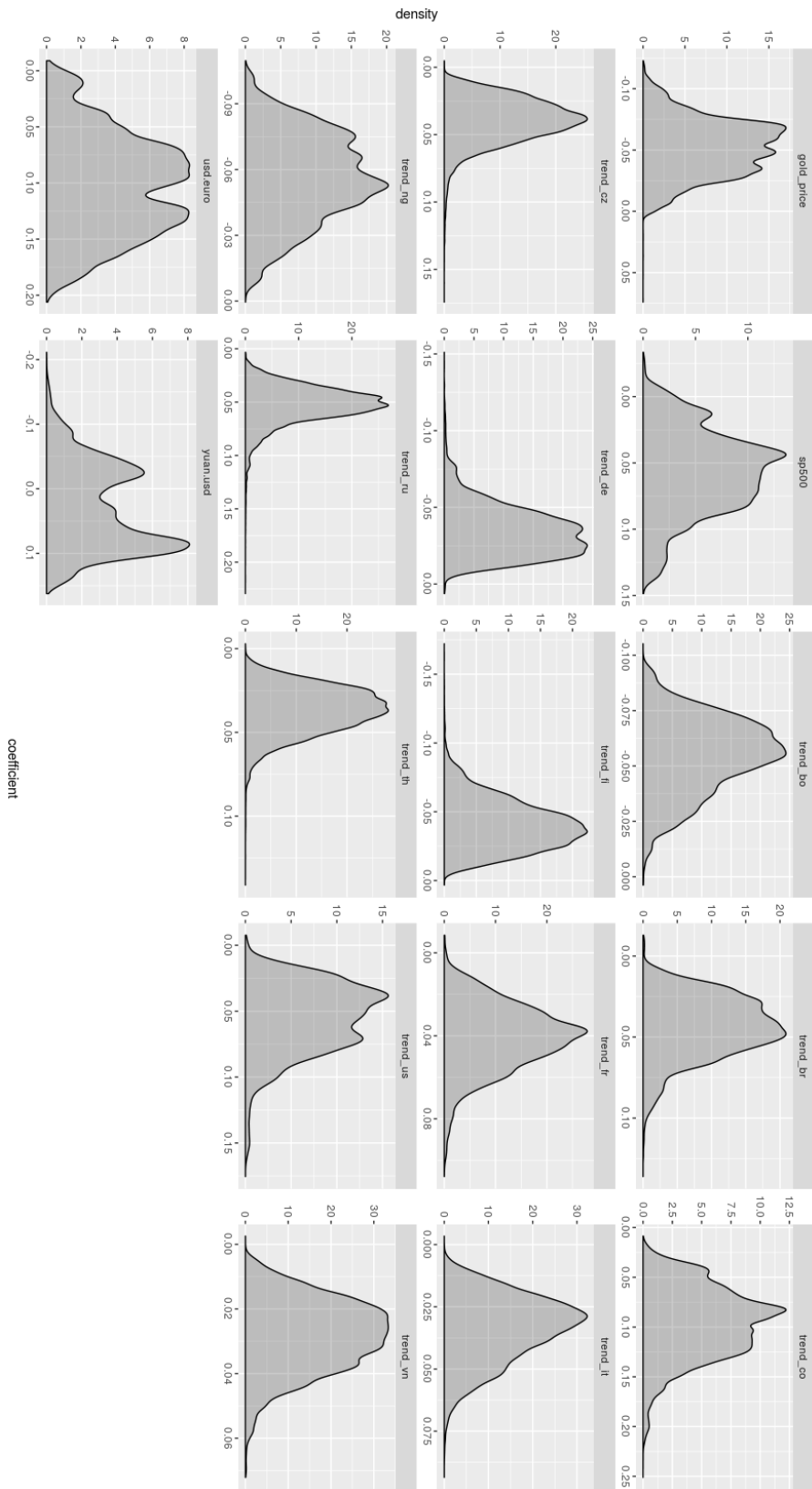

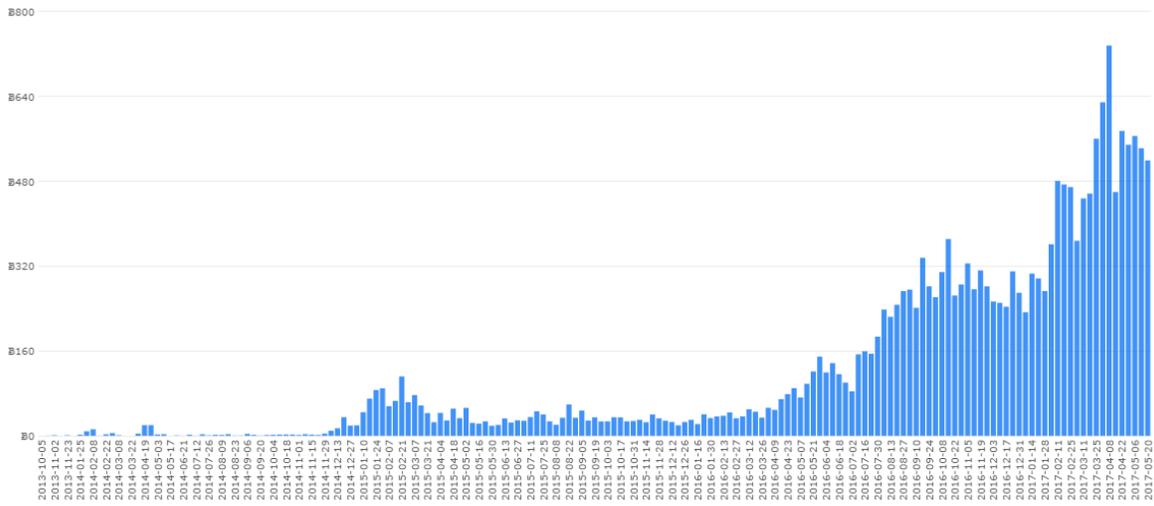

Figure 13: Weekly Localbitcoins Volume